\documentclass[aps,prl,twocolumn,showpacs,preprintnumbers,amsmath,amssymb]{revtex4}

\usepackage{graphicx}
\usepackage{dcolumn}
\usepackage{bm}
\begin{document}

\title{Effect of periodic parametric excitation on an ensemble of force--coupled
self--oscillators}

\author{E. Y. Shchekinova$^1$}
 \email{elenash@pks.mpg.de}
\affiliation{
 \\$^1$Max Planck Institute for the Physics of Complex Systems, N\"{o}thnitzer Str 38, 01187 Dresden, Germany
}

\date{\today}
\begin{abstract}

We report the synchronization behavior in a one--dimensional chain of
identical limit cycle oscillators coupled to a mass--spring load via
a force relation. We consider the effect of periodic parametric
modulation on the final synchronization states of the system. Two
types of external parametric excitations are investigated
numerically: periodic modulation of the stiffness of the inertial
oscillator and periodic excitation of the frequency of the
self--oscillatory element. We show that the synchronization
scenarios are ruled not only by the choice of parameters of the
excitation force but depend on the initial collective state in the
ensemble. We give detailed analysis of entrainment behavior for
initially homogeneous and inhomogeneous states. Among other results, we describe a regime of partial
synchronization. This regime is characterized by the frequency of
collective oscillation being entrained to the stimulation frequency
but different from the average individual oscillators frequency.



\end{abstract}
 \maketitle

\section{Introduction}\label{sec.1}

In Nature many life significant processes are regulated by mechanisms of
synchronization and entrainment. A properly chosen external stimulus
can synchronize an intrinsic biological cycle. A great number of
biological processes involve several oscillatory cycles that
interact and contribute to an overall system temporal behavior. To
give a few examples, the human heart responds sensitively to the forced
oscillations modulated by a pacemaker with various
frequencies~\cite{glass1991}, the respiratory rhythm in humans and
animals can be entrained to a mechanical ventilator
phase~\cite{graves1986}. Fibrillar flight muscles of some insects
act as a mechanical resonator that is coupled to the insects thorax
and wings~\cite{dickinson2006}. A rhythm of flight muscles
contractions and wings vibration frequency can be turned by an
experimental mechanical or optical
stimulation~\cite{nalbach1988,lehmann1997}. In all the examples
provided so far a self--oscillatory rhythm is modified due to
mechanical forces via the coupling to an external mechanical
oscillator. A simple theoretical description in terms of interacting
nonlinear oscillators would provide a universal approach for
exploiting dynamics in this type of systems. 

For analysis of time evolution processes simplified models from
nonlinear dynamics have been used~\cite{pikovsky2001}. In fact, in
the framework of nonlinear models one can describe the phenomena
that occur in a relatively large class of biological and chemical
systems~\cite{glass2001,abarbanel2006}. In this paper we use the
analogies found in nature and introduce a non--feedback model of
entrainment in a system of coupled nonlinear oscillators. Our 
model might help to elucidate the basic underlying
mechanisms that regulate synchronization of processes in a real
biosystem.

 Although simplified
low--dimensional models can capture inherent mechanisms leading to
synchronization, quite often it happens that spatial dynamics plays
a crucial role in mediating synchronization~\cite{chate1999} in a
given system. To unveil the whole variety of possible
synchronization scenarios one has to introduce a spatial dimension
in the mathematical model of a biosystem.

Collective phenomena in spatially extended systems and their
one--dimensional chain analogues have been a subject of a large body
of investigations~\cite{pikovsky2001,afraimovich1994,osipov2007}.
  For instance, models of self--oscillatory systems
are widely used to describe biochemical pattern formation
processes in spatially extended
systems~\cite{battogtokh1996,rudiger2007}. It is well known that the
complex spatiotemporal behavior in a system is largely determined by
its dynamical instabilities. There are two basic approaches that are
used for numerical and analytical studies of stability in spatially
extended systems subjected to external forcing. The first approach
implements the mathematical framework provided by the complex
Ginzburg--Landau equation (CGLE)~\cite{kuramoto1984}. In this case
the description of an oscillatory medium is reduced to the study of
the amplitude equation with a periodic forcing. Different types of
periodic forcing of the CGLE and a rich set of dynamical states and
their bifurcations have been studied
in~\cite{chate1999,battogtokh1996,coullet1992,rudiger2007}. The
second approach uses a description of system spatiotemporal
dynamics in terms of a large populations of coupled oscillators. Due
to a large number of degrees of freedom these systems can display a
panoply of dynamical behaviors: from cluster formation
~\cite{drendel1984,hakim1992,vadivasova2001} to multistability
regimes~\cite{topaj2002,osipov2007} with the coexistence of distinct
stable collective modes of oscillations. 

Most of the theoretical
studies mentioned so far are devoted to investigation of dynamics in
globally or locally coupled ensembles of oscillators. The coupling
term usually depends on phases or displacements of oscillators.
In this work we discuss a new type of coupling via a common
force. This force explicitly depends on the displacements as
well as on the velocities and the accelerations of individual elements in a large
ensemble. The idea to introduce such a coupling has been first
proposed in Ref.~\cite{vilfan2003} The model of a large ensemble of
self--oscillatory elements coupled via a common force to an inertial
oscillator was proposed in~\cite{vilfan2003} as a phenomenological
approach to study spontaneous oscillations in an active muscle fiber
coupled to mechanical resonator~\cite{yasuda1996}. Each element in
the ensemble is described by the Stuart--Landau equation with
real coefficient. 
It is shown that the system undergoes a Hopf bifurcation near to a
critical threshold defined in systems parameter space. Various
regimes of collective oscillations are reported to exist. In fact,
the systems parameter space is divided into several regions with
mono-- and multi--stable collective modes of oscillatory dynamics.
Near to an instability threshold defined in the parameter space there exist completely synchronous, asynchronous
and antisynchronous regimes of collective dynamics. In addition,
regions with the coexistence of several modes are found.

In this paper we introduce a modification of the model considered in
Ref.~\cite{vilfan2003}. We include an external time--dependent
stimulation by adding small amplitude time--dependent
terms in the parameters expression. We describe the
mechanism of entrainment of the frequency of system solutions to the stimulation frequency. 
The essential feature of the
described non--feedback mechanism of entrainment is that the
external stimulation enters in the model as a parametric modulation.
Two types of parametric excitations are treated: the first type
is the periodic modulation of the inertial oscillator stiffness;
the second type is the peridic excitation of a natural frequency of a
self--oscillatory element.  We show that the synchronization
scenarios are ruled not only by the choice of parameters of the
excitation force but depend on the initial collective state in the
ensemble.  For both
types of the periodic excitations entrainment behavior is
studied for homogeneous states. We find sets of
frequency--locked solutions that correspond to the resonant driving.
We also study the effect of two types of periodic driving on stability of inhomogeneous states.
In this work we address several questions: what is the influence
of periodic parametric forcing on the stability of a particular
collective motion, and how does the features of synchronization compare
for different types of parametric excitations in the system.

This paper is organized as follows. In Section~\ref{sec.2}, we
introduce the autonomous model and briefly recall some features of
the complex phase dynamics from~\cite{vilfan2003}. Next, in
Section~\ref{sec.3} we provide the time--dependent model
with the first type of parametric modulation and investigate the
existence of stable periodic solutions for the case of $N=1$ oscillator in
the chain. We explain the numerical procedure that is used for the
amplitude--frequency response calculations. Amplitude--frequency responses from numerical
simulations and analytical derivation are compared. Next, we present, using numerical
simulations, the effect of the periodic forcing on initially
inhomogeneous states behavior. We list a variety of regimes of inhomogeneous
collective dynamics including partially synchronized states. The
second type of parametric modulation is introduced in
Section~\ref{sec.4}. Similarly to the previous section, we provide
stability diagrams for the case of $N=1$ oscillatory unit in the chain coupled with an
inertial element. We discuss a biological relevance of our findings and give
general conclusions in Section~\ref{sec.5}. In the Appendix an analytical amplitude--frequency
 relation is derived using a first
order quasiperiodic approach .
\section{Model equations}
\label{sec.2}

In this section we review the basic model \cite{vilfan2003} of a
chain of active mechanical oscillators coupled to a damped
mass--spring oscillator.
  Each elementary unit in the chain can be modeled by the Stuart--Landau
equation. Two variables are defined describing the amplitude
$\mathrm{r}_i$ and the phase $\phi_i$ of each $i$th element in a
finite chain of $N$ elements:
\begin{equation}
\dot{\mathrm{r}}_i=\epsilon\mathrm{r}_i-B\mathrm{r}_i^3,\qquad \dot{\phi}_i=\omega,\label{eq.1}
\end{equation}
where $B$ is the Landau coefficient. The set of $N$ equations for
$i=1,\ldots N$ describes $N$ uncoupled oscillators. In this paper we
consider mechanical elements all arranged in a one--dimensional
chain and coupled via a common force due to a linear inertial
oscillator attached to one end of the chain. In the
model~\cite{vilfan2003} a chain of $N$ elements has one fixed
boundary and is connected to a mass load in a way that the change
of the total extension of $N+1$ oscillators is zero. This corresponds to the no flux
boundary condition. Consequently, this condition can be recast in the form of
the following geometrical constraint imposed on oscillators
motion:
\begin{equation}
\sum_{i=1}^{N+1} x_i=0,\label{eq.2}
\end{equation}
Where $x_i=\mathrm{r}_i \cos \phi_i, i=1,\ldots N$ is the
displacement of the $i$th active self--oscillatory element,
$x_{N+1}$ is a displacement of the linear inertial oscillator. The
external force due to the mass--spring load is determined by the
equation of motion:
\begin{equation}
F=-M\sum_{i=1}^{N}(\ddot{x}_i+\Omega_0^2x_i+\nu
\dot{x}_i)\label{eq.3}
\end{equation}
where the parameters $M,\Omega_0$ and $\nu$ are the mass,
characteristic frequency of linear oscillator and the spring
damping. The spring stiffness coefficient is $k=M\Omega_0^2$. Upon
including the mechanical force $F$ from (\ref{eq.3})
Eqs.(\ref{eq.1}) can be written in the complex
form~\cite{vilfan2003}:
\begin{equation}
\dot{z}_i=(\mathrm{i}\omega+\epsilon)
z_i-B|z_i|^2z_i+F\xi^{-1},\label{eq.4}
\end{equation}
where $z_i=r_i e^{\mathrm{i} \phi_i}$ and $\xi$ is the scaling parameter. The set
of equations (\ref{eq.3})-(\ref{eq.4}) represents a system of active oscillators that are
coupled via a mean field $F/\xi$. Unlike in the Kuramoto model of
phase oscillators with global coupling ~\cite{kuramoto1984} here the
collective behavior is governed by a mean field that depends on
the velocities and the accelerations of elements in the chain. It was shown
in Ref.\cite{vilfan2003} that oscillatory versus non--oscillatory
motions exist in an unfolded space of the system parameters. Namely,
oscillatory versus non--oscillatory behavior depends on the values
of stiffness, mass of inertial load and the bifurcation parameter
$\epsilon$. In fact the onset of the oscillatory behavior happens
for $\epsilon<0$ when the system undergoes a Hopf bifurcation at the
critical value of linear frequency $\Omega_0=\Omega_c$. The critical
frequency for which transition from a stable non--oscillatory to an
oscillating behavior occurs is defined in~\cite{vilfan2003} from the
expression: $\Omega_c=\{\omega^2/(1+N M
\nu/\xi)+\epsilon(\nu-\epsilon+2 \xi/(N M))\}^{1/2}$. While 
oscillatory states can be found for $\Omega_0<\Omega_c$, for higher
values of the linear frequency $\Omega_0>\Omega_c$ initial oscillations
always decay to the stable trivial solution.

 A rich bifurcation diagram
with various modes of collective oscillatory dynamics exists for
$\epsilon>0$ near the critical instability
threshold~\cite{vilfan2003}. Several regions with homogeneous and
inhomogeneous collective modes as well as regions with
coexistence of different dynamical regimes are found. Depending on
the frequency $\Omega_0$ and $\epsilon$ one can observe different
asymptotic states of system behavior: asynchronous, antisynchronous
and synchronous states. For a low value of the stifness $\Omega_0\ll\Omega_c$ the system
is set into a synchronized oscillation state. The phase and the amplitude of
self--oscillatory elements in a globally synchronized regime are
identical.  For large positive $\epsilon$ and for
$\Omega_0$ nearly close to the instability threshold value
$\Omega_c$ a set of asynchronous states is found. The asynchronous state can be characterized by an
inhomogeneous distribution of phases and amplitudes of oscillators
along the chain. Remarkably, that while the total chain extension vibrates at one
frequency the value of the average oscillators frequency is below it~\cite{vilfan2003}. 
Localized regions of synchronized motion can be identified along the chain.  For large value of the
frequency $\Omega_0$ and $\epsilon>0$ the system is in an
antisynchronous regime. In this state neighboring oscillators move in an antiphase so that the
total chain extension remains fixed.

\begin{figure}
\begin{center}
\includegraphics[width=8.cm]{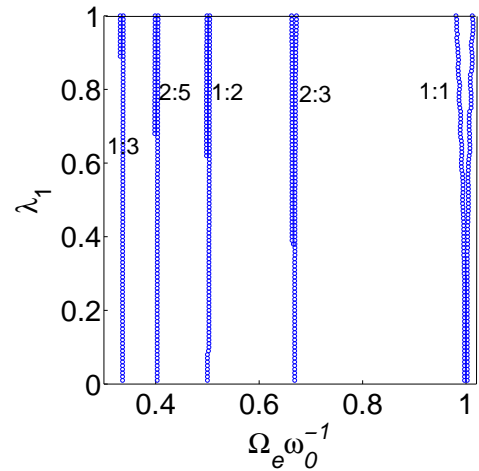}
\caption{\label{fig:fig1} Arnol'd tongues: stability diagram for the
parameter space defined by the amplitude $\lambda_1$ and frequency
ratio $\Omega_e\omega_0^{-1}$. The boundaries that separate stable limit
cycles from unstable and quasiperiodic solutions are shown. Frequency ratios of entrained limit cycle
solutions inside each stability region are indicated. Parameters:~$\epsilon=-0.2,~\nu=0.2,\xi=0.3125,B=1,\omega=1,M=1,k=0.36 $.
Original unperturbed limit cycle frequency:~$\omega_0=0.7439$.}
\end{center}
\end{figure}

\begin{figure}
\begin{center}
\includegraphics[width=8.cm]{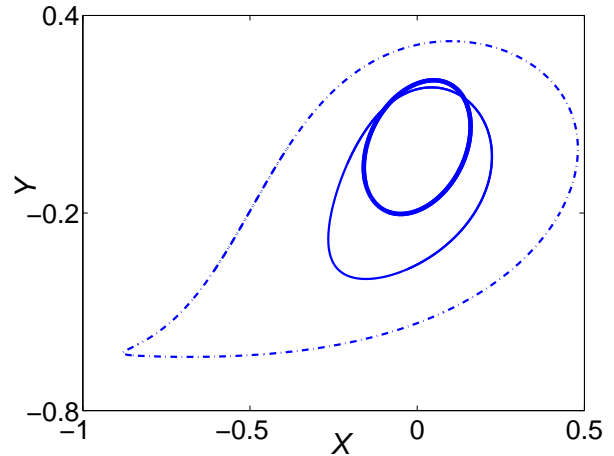}
\caption{\label{fig:fig2} Examples of limit cycle solutions. Thick
line shows the original limit cycle for $K(\lambda_1,t)=k$
in system~(\ref{eq.4})-(\ref{eq.6}). Two other limit cycle solutions
are calculated for the system~(\ref{eq.4})-(\ref{eq.6}) with
non--zero amplitude of driving $\lambda_1=0.5~k$ (thin line), and
with $\lambda_1=k$ (dash--dotted line). External frequency
$\Omega_e=\omega_0=0.7439$ and the stiffness coefficient $k=0.36$. }
\end{center}
\end{figure}
\section{Effect of periodic modulation of
stiffness}\label{sec.3} In this
section, we discuss the effect of a non--additive periodic
excitation on lock--in solutions stability in the system of self--oscillators coupled
to an inertial load.  The excitation is taken in the form of
periodic modulation of the inertial oscillator stiffness coefficient.
First we define a set of equations of motion and then proceed to examining the system stability domains.
We study the amplitude dependence for frequencies of excitation near to the one--to--one resonance.
Let us introduce a parametric modulation in the stiffness $k$ of the inertial
oscillator~(\ref{eq.3}) by using a periodic term
$q(\lambda_1,t)=\lambda_1 \cos \Omega_e t$ with frequency
 $\Omega_e$ and amplitude $\lambda_1$. Then the expression for the force is:
\begin{equation}
F=-M \sum_{i=1}^N\mathrm{Re}(\ddot{z}_i+K(\lambda_1,t)
z_i+\nu\dot{z}_i)\label{eq.5}
\end{equation}
where $K(\lambda_1,t)=k+q(\lambda_1,t)$ is the time--dependent
stiffness coefficient. In our treatment we include the case of a strong
magnitude of driving by taking $\lambda_1/k \leq 1$. Upon
differentiating both side of Eq.~(\ref{eq.4}) and substituting the
expression for $\ddot{z}_i$ from~(\ref{eq.5}) the differential
equation for the force yields:
\begin{eqnarray}
\dot{F}=-\frac{\xi F}{N
M}+\frac{\xi}{N}\sum_{i=1}^N\mathrm{Re}\bigg\{\frac{d \left( B\vert
z_i \vert^2
z_i+\mathrm{i} \omega z_i\right)}{dt}\nonumber\\
-(\epsilon+\nu)
\dot{z}_i-K(\lambda_1,t) z_i\bigg\},\label{eq.6}
\end{eqnarray}
Together with Eq.~(\ref{eq.4}) we now obtain a complete set of differential
equations for $N+1$ variables. The displacement $x_{N+1}$ of the inertial oscillator
is determined from the geometrical constraint~(\ref{eq.2}).
We expect that the stability behavior in system (\ref{eq.4})-(\ref{eq.6}) 
is largely determined by a new type of coupling. 
As a consequence of the coupling the motion of every self--oscillatory
element is affected by a common force. This force is explicitly dependent on time as well as on the oscillators
displacement, velocity and acceleration.
\subsection{Entrainment for an initially synchronized regime}
\label{sec.3.1}

Let us first discuss the entrainment behavior for initially
homogeneous solutions: all the oscillators in ensemble are moving in
phase and with the same frequency of oscillations. The linear
stability analysis of the time--independent system in
Ref.~\cite{vilfan2003} provides a detailed phase diagram with the
regions of stable globally synchronized solutions. Indeed, for the
parameters space corresponding to $\epsilon<0$ and
$\Omega_0<\Omega_c$ there exists a large set of stable synchronized
solutions of Eqs.~(\ref{eq.4})-(\ref{eq.6}) for $\lambda_1=0$. When
the synchronized state is stable it suffices to reduce systems
description to the case of $N=1$ oscillator coupled with inertial
load. However, when the parametric modulation is introduced in the model the
stability problems for the system of $N=1$ and $N>1$ oscillators in the chain are no longer
equivalent. The applied periodic excitation can change the
stability of the synchronous solution for $N>1$.  Thefore, the system can
be driven to a new asymptoticcally inhomogeneous state. In our work we do not analyze
the stability of synchronized solutions for the entire parameter
space. Instead, we consider entrainment behavior of the synchronized
state for a particular choice of parameters: $\epsilon=-0.2$ and
$\Omega_0=0.6$. We integrate numerically the time-dependent
system~(\ref{eq.4})-(\ref{eq.6}) with $N=1$ and $N=100$ oscillators. It follows that the observed entrainment behavior for the 
low--$(N=1)$ and high--dimensional $(N>1)$ cases matches well for the choice of
parameters given above. Therefore, in order to simplify our analytical
derivations and to speed up numerical procedure we reduce dynamical
description of the system to the low--dimensional case. In the following
paragraphs we discuss entrainment behavior for $N=1$
oscillator coupled with a mass load.
For initial conditions we chose
arbitrary small displacement $x_1=\mathrm{z_1}$. Results
of numerical simulation are given in Fig.~(\ref{fig:fig1}). We show
stability domains in the parameter space defined by the amplitude of
driving $\lambda_1$ and the ratio between external frequency
$\Omega_e$ and the original limit cycle frequency $\omega_0$. Here
the stability regions with lock--in solutions are separated from the
regions of quasiperiodic and unstable solutions by defined
boundaries. Inside each stability region the lock--in solution has
rational frequencies ratio $m:n$. Note that the width of entrainment
regions shortens as one proceeds from the $1:1$ resonance to the
lower frequency ratios. Several limit cycle solutions are presented in
Fig.~(\ref{fig:fig2}) for the $1:1$ resonance and for different amplitudes $\lambda_1$.

\subsection{Amplitude--frequency response} \label{sec.3.2}

In this section we present numerical and analytical results on the
amplitude--frequency dependence for the system~(\ref{eq.4})-(\ref{eq.6}).
 We use numerical analysis of fundamental
frequencies~\cite{laskar1992} to determine the amplitude of 
quasiperiodic and periodic solutions of the
system~(\ref{eq.4})-(\ref{eq.6}) for the frequencies close to the one--to--one stimulation.
 For solutions $x=\mathrm{Re} z_1$ and $y=\mathrm{Im} z_1$ of~(\ref{eq.4})-(\ref{eq.6}) we obtain a quasiperiodic
approximation near the vicinity of the original limit cycle solution
of Eqs.~(\ref{eq.4})-(\ref{eq.6}) for $\lambda_1=0$.
 By using the substitution of solutions $x(t)$ and $y(t)$ with
their zero time--averaged equivalents: $x(t)\rightarrow
x(t)-\bar{x}$~and~$y(t)\rightarrow y(t)-\bar{y}$
 we write down the quasiperiodic expansions for the new $x(t)$ and $y(t)$:
\begin{equation}
x(t)=\sum_{q}c_q e^{\mathrm{i} \gamma_q t},~\quad~y(t)=\sum_{q}d_q
e^{\mathrm{i} \gamma_q t},\label{eq.7}
\end{equation}
where ${\gamma_q}$ is a set of time--independent frequencies. We assume 
that in the above expression the amplitudes ${c_q}$ and ${d_q}$ do not dependent on
time.
In the above expressions the terms are collected according to
decreasing order of magnitude. To begin our numerical procedure we
 introduce a set of new coefficients $\tilde{c}_n$ and
$\tilde{d}_n$ that are expressed from the following
integrals~\cite{laskar1992}:
\begin{subequations}
\begin{eqnarray}
\tilde{c}_n&=&\frac{1}{2 T}\int_{-T}^{T} x(t) e^{-\mathrm{i} n (\pi/T)t} \chi(t) dt,\\
\tilde{d}_n&=&\frac{1}{2 T}\int_{-T}^{T} y(t) e^{-\mathrm{i} n
(\pi/T)t} \chi(t) dt,\label{eq.8}
\end{eqnarray}
\end{subequations}
where $\chi(t)=1+\cos \pi t/T$ is the Hanning window. The interval
of time $100 \pi\leq T \leq 300\pi$ is the integration time for every systems
trajectory. Next,
 two sequences of coefficients ${c_q}$,~and
${d_q}$ are produced by rearranging
$\bar{c}_n$ and $\bar{d}_n$
according to the decreasing order of magnitude. In the expansions ~(\ref{eq.7})
we only retain the first two terms that produce reasonably accurate
approximation to the solutions $x(t)$ and $y(t)$. We calculate the first
four largest coefficients of the series ${c_q}$~and
${d_q}$ as follows:
\begin{subequations}
\label{eq.9}
\begin{eqnarray}
c_{1,-1}=\max\lbrace{\tilde{c}_{1,-1},\tilde{c}_{2,-2},\ldots
\rbrace},\\d_{1,-1}=\max\lbrace{\tilde{d}_{1,-1},\tilde{d}_{2,-2},\ldots
\rbrace},
\end{eqnarray}
\end{subequations}
\begin{figure}
\begin{center}
 \includegraphics[width=8.cm]{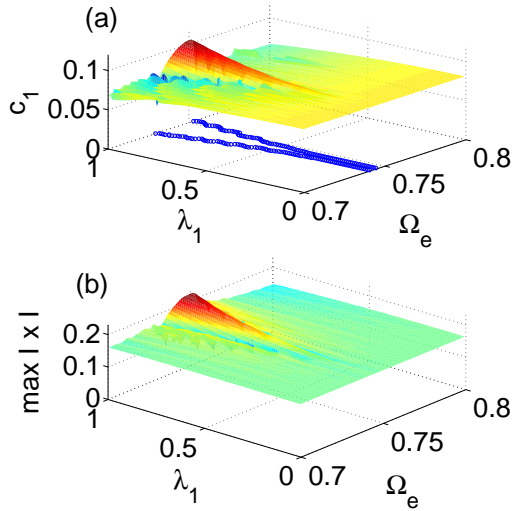}
 \caption{\label{fig:fig3} Amplitude--frequency response surface is shown
on the $(\Omega_e,\lambda_1)$ plane. (a) Numerically calculated
amplitude $c_1$ of the leading frequency component of the limit cycle in
Eq.~(\ref{eq.5}) near to the $1:1$ resonance for the driven system
(\ref{eq.4})-(\ref{eq.6}). Stability boundaries of the $1:1$ Arnol'd
tongue are shown on the $(\Omega_e, \lambda_1)$ plane. (b) Numerically
estimated maximum of the limit cycle solution for
(\ref{eq.4})-(\ref{eq.6}) with $\max|x|=\max_{{200 \pi\leqslant t\leqslant
300 \pi}} |x(t)-\bar{x}|$ for different amplitudes $\lambda_1$ and
frequencies $\Omega_e$ of external driving.
 Parameters are: $\epsilon=-0.2,\nu=0.2,k=0.36, M=1,\xi=0.3125$.}
 \end{center}
 \end{figure}
\begin{figure}
\begin{center}
\includegraphics[width=8.cm]{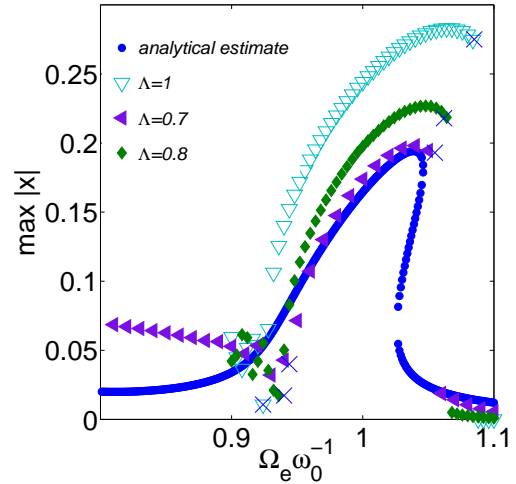}
\caption{\label{fig:fig4} Amplitude--frequency response curves
obtained at a fixed amplitude of $\Lambda=\lambda_1/k=[0.7,0.8,1]$
from Fig.~(\ref{fig:fig3}). Analytical amplitude--frequency
dependence $\Gamma_{\tilde{\lambda}_1}(\Omega_e)$ is displayed (small
circles) (see appendix for derivation). Here $\Omega_e
\omega_0^{-1}=\Omega \Omega_0^{-1}(1-\Delta)$, where
$\Delta=0.07\Omega^{-1}$ is the frequency mismatch. Parameters are
given as in the Appendix. Numerically evaluated stability boundaries
are shown (crosses) for all three values of $\Lambda$. Parameters
for the analytical amplitude--frequency dependence:
$\epsilon=-0.2,\nu=0.2,M=1,\omega=1,\xi=0.3125,\Omega_0=0.7861,\lambda_1=M
\Omega_0^2$}
\end{center}
\end{figure}

 Figure~(\ref{fig:fig3}a) illustrates the parametric dependence of the largest
coefficient $c_1$ in the expansion~(\ref{eq.9}a)
 on the amplitude $\lambda_1$ and the frequency $\Omega_e$. Numerical responses are obtained by
taking initial conditions for $x(t)$ and $y(t)$ to be randomly
distributed in the vicinity of unperturbed limit cycle for
Eqs.~(\ref{eq.4})-(\ref{eq.6}). For comparison we also calculated the maximum of the numerical
solution over the same time interval $T$.
Figure~(\ref{fig:fig3}b) shows numerical results for the
maximum of the amplitude $x(t)$:~ $\max|x|=\max_{T} |x(t)|$. 
In the plane $(\Omega_e, \lambda_1)$ stability boundaries for the $1:1$ resonance are indicated. 
The results from two plots are consistent except for the regions near to the stability
borders. The
amplitude--frequency response is a surface that is smooth inside the
borders of stability region and shows discontinuities outside near
to the borders of Arnold tongue. Apparently, the discontinuities are
due to the emergence of instabilities.
Because of the instabilities, the leading frequency components that
are defined from the quasiperiodic approximation (\ref{eq.7}) in
this case might be distinctly different. Consequently, the numerical
procedure produces a set of amplitudes which are not smoothly
dependent on the external frequency $\Omega_e$. 

We compare
numerically estimated amplitude--frequency responses with the
results obtained by using quasiperiodic approximation (see
Appendix).
 In Fig.~({\ref{fig:fig4}}) three numerically calculated
amplitude--frequency curves are displayed for different values of
the amplitude $\Lambda=\lambda_1/k$. To compare these results with
the analytical calculations, we display the amplitude--frequency curve
$\Gamma_{\tilde{\lambda}_1}(\Omega_e)$ obtained from the quasiperiodic
approximation of limit cycle solution $(x(t),y(t))$ (see
Eq.~(\ref{eq.17}) in the Appendix). It is apparent, that for a
given choice of parameters the curves from numerical and
analytical calculations follow similar nonlinear response behavior
inside the stability region (stability boundary is marked by
crosses).

\subsection{Entrainment for anti- and asynchronous
regimes}\label{sec.3.3} In this section we illustrate results of
numerical analysis for the periodic forcing of initially spatially
inhomogeneous states. The existence and stability of these states
for the autonomous system ($\lambda_1=0$) is controlled by the
choice of a mass of inertial oscillator and parameter $\epsilon$.
For the time--dependent system it is expected that under the
influence of a strong driving these states might become unstable and
converge to a new asymptotic state. In the
limit of a large excitation it can produce significant changes on
the regions of stability of existing spatiotemporal regimes. We demonstrate that,
meanwhile some collective regimes do not survive new spatially
homogeneous states become asymptotically stable. Here we show which
conditions on the frequency and the amplitude of external driving have
to be satisfied in order for the resulting spatial state to be
stable and homogeneous.
\begin{figure}
\begin{center}
\includegraphics[width=8.cm]{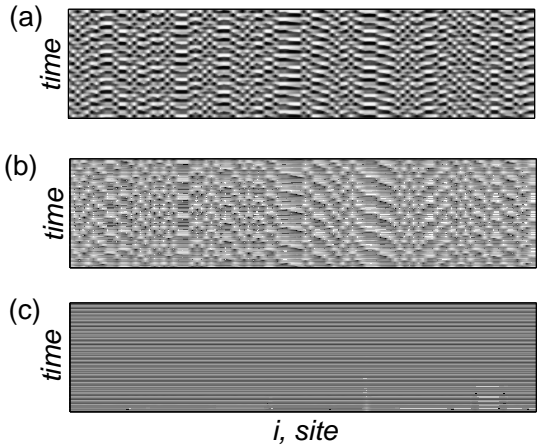}
\caption{\label{fig:fig5} Space--time diagrams of intensity of
oscillations $|z_i|$ for the $i$th element. (a)  Asynchronous state
for time--independent system~
(\ref{eq.4})-(\ref{eq.6})~($\lambda_1=0$). (b) Asynchronous state
for the time--dependent
system~(\ref{eq.4})-(\ref{eq.6})~($\lambda_1=k$).The frequency of
driving $\Omega_e=\omega_0=0.85$. .(c) Synchronized behavior is
obtained with the parametric forcing of an initially asynchronous
state of system~(\ref{eq.10})-(\ref{eq.11})~($\lambda_2=0.5
\omega, \Omega_e=0.85$). Parameters for (a),(b) and (c):
$\Omega_0=0.85,\epsilon=0.5, k=M \Omega_0^2,
M=1,\omega=1,\xi/N=0.3125, N=100$. }
\end{center}
\end{figure}
\begin{figure}
\begin{center}
\includegraphics[width=8.cm]{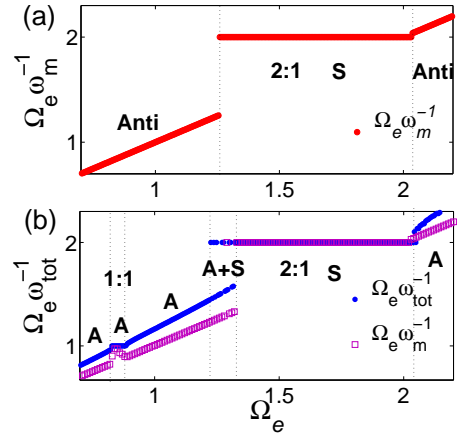}
\caption{\label{fig:fig6} Frequency ratio curves for the
antisynchronous (Anti) (asynchronous (A)) state of a chain of $N=100$
oscillators. (a) Ratio between external frequency
 $\Omega_e$ and the average oscillator frequency $\omega_m=1/N\sum_i^N \omega_i$ is plotted
 versus external frequency for an initially antisynchronous state of
the system~(\ref{eq.4})-(\ref{eq.6}).(b) Frequency curves are plotted for an
initially asynchronous state of the system
~(\ref{eq.4})-(\ref{eq.6}). Shown are two frequency ratio
curves: $\Omega_e\omega_m^{-1}$ versus $\Omega_e$ (squares) and
 $\Omega_e\omega_{tot}^{-1}$ versus $\Omega_e$ (dots), where $\omega_{tot}$ is the frequency of the
total extension $x_{tot}$ of a chain of oscillators.
Parameters: $\epsilon =0.1$ for antisynchronous state, ($\epsilon=0.5$) for asynchronous state, $\Omega_0=0.85, k=M \Omega_0^2, M=1, \omega=1,\xi/N=0.3125,\nu=0.2$ and $\lambda_1=k$ for both (a) and (b).
The regions of stable synchronized states are indicated by (S).}
\end{center}
\end{figure}
 In the literature studies of collective dynamics of chains of
coupled oscillators are often reduced to their phase dynamics
studies~\cite{daido1996,pikovsky2001}. Our system is different from
these cases because the amplitude of external force $F$ strongly
depends on the state of the $i$th oscillator, and therefore, on the
collective dynamics in an ensemble. It is not sufficient
to confine the investigation only to a phase dynamics study, like in
the cases of weakly coupled oscillators \cite{ermentrout1991}. It
should be pointed out, that in the limit of a strong excitation the
effect on amplitude variation of each individual oscillator has to
be taken into account. Namely, for a large relative modulation
$\lambda_1/k\sim 1$ not only the individual oscillator frequency is
affected but also the amplitudes.

First, we proceed by considering the spatiotemporal state of the
system ~(\ref{eq.4})-(\ref{eq.6}) with and without external
parametric excitation. The time evolution of $N=100$
self--oscillators from an initially inhomogeneous collective state
is followed by plotting space--time diagrams. These diagrams display
a dynamical collective state of the system. In order to do so, we
integrate $2N+1$ equations. The initial conditions are prepared by
adding random perturbations to the initial zero state.
Figure~(\ref{fig:fig5}) shows the space--time diagram of the
amplitude of $i$th oscillator $|z_i|$ on the $(i,t)$ plane for
$N=50$ oscillators. The oscillator number is shown in abscissa and
time $t$ in ordinate. Figure~(\ref{fig:fig5}a) displays the time
evolution of an asynchronous state from the initial time $t=0$ to
the final time $t=300 \pi$ in the absence of external periodic
driving. A closer look reveals that the elements in the half of the
ensemble ($i=1,\ldots 50$) are grouped in small clusters of
oscillators. Each group consists of elements that are phase shifted
with respect to the neighboring elements of the same group. As the
external stimulation turned on the state in Fig.~(\ref{fig:fig5}a)
evolves to a new asynchronous state shown in the panel (b) of the
same figure. It presents the time--evolution of an inhomogeneous
unstable state for $N=50$ oscillators and the frequency of
excitation $\Omega_e=0.85$.
 It can be easily seen that the distribution of groups of oscillators remains the same as
in Fig.~(\ref{fig:fig5}a) but the average self--oscillator
frequency is shifted towards the stimulation frequency $\Omega_e$.
The state is not stable, and is sensitive to variations of initial
distributions of the positions and the velocities of the oscillators. In
fact, the value of numerically estimated Lyapunov exponent is
positive in this case.  Such regimes are found for the CGLE with an
additive periodic forcing~\cite{chate1999}. These "turbulent
synchronized states" arise in the situation when the forcing is weak
and its frequency is chosen nearly the same as a natural oscillation
frequency~\cite{chate1999}. In this case the dynamical state is
characterized by the average frequency of collective excited
oscillations being equal to external frequency and its numerical
Lyapunov exponent being greater than zero.
\begin{figure}
\begin{center}
\includegraphics[width=8.cm]{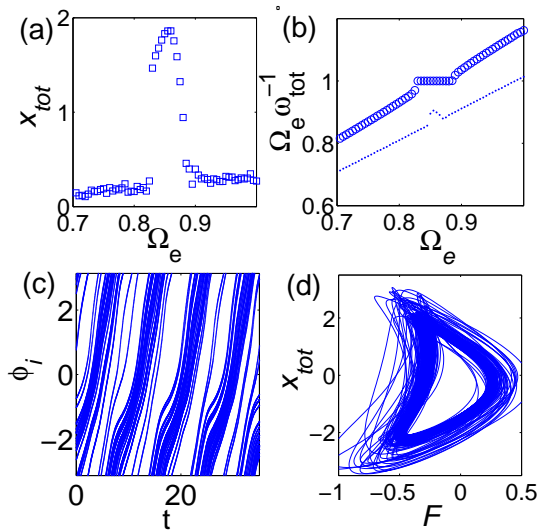}
\caption{\label{fig:fig7} Asynchronous state of oscillators near
the $1:1$ resonant driving region from Fig.~(\ref{fig:fig6}b).~(a)
Amplitude of total extension $x_{tot}=\sum_{i=1}^{N} x_i$ of a chain
of $N=100$ oscillators versus external frequency $\Omega_e$.~(b) Two
frequency ratios versus external frequency are displayed:
$\Omega_e\omega_m^{-1}$ (dots) and $\Omega_e\omega_{tot}^{-1}$
(circles). $\omega_m$ is the average frequency of oscillators in a
chain, $\omega_{tot}$ in the frequency of total extension.~(c) Phase
$\phi_i=\arg(z_i)$ of $N=100$ active elements as a function of time
for asynchronous state solution for $\Omega_e=0.85$.~(d) Example of
quasiperiodic solution plotted for $x_{tot}$ versus
force $F$.~$\Omega_e=0.85$ for~(c) and (d).}
\end{center}
\end{figure}
Finally, a perfectly synchronized solution is displayed in
Fig.~(\ref{fig:fig5}c). Unlike in the previous two plots, here
after a short transient time the ensemble is organized into coherent
structure with all the individual elements moving in phase and with
the frequency $\Omega_e$. To produce this plot another type of
stimulation is used that will be discussed in the following
sections.

 We will specify the transitions that occur in
the system for various stimulation frequencies by plotting the
frequency ratio curves versus the stimulation frequency.
Figure~(\ref{fig:fig6}) displays two cases for distinct parameter
choices. Frequency ratio curves for an initially antisynchronous state
and for an initially asynchronous state are shown in
Fig.~(\ref{fig:fig6}) (a) and (b).  Let us first examine the changes in
dynamical behavior if we chose the parameters from the domain that
corresponds to an antisynchronous oscillatory regime stable for the
time--independent system. This can be easily done by defining the
bifurcation parameter $\epsilon=0.1$ and by taking the frequency of
the inertial oscillator $\Omega_0=0.85$~\cite{vilfan2003}. We plot
in Fig.~(\ref{fig:fig6}a) the frequency ratio curves
$\Omega_e\omega_m^{-1}$ versus $\Omega_e $. Several distinct regions
that correspond to different collective dynamical states exist. The
system remains in the antisynchronous state for an interval
$\Omega_e\in[0.2;1.2]$. For higher frequencies
$\Omega_e\in[1.22;2.03]$ the solution is driven into the synchronized
$2:1$ state. This synchronized state is characterized by all the
oscillatory elements moving in--phase and with the same frequency.
Upon increasing the excitation frequency $\Omega_e\geq2.03$ the
system undergoes a transition to an antisynchronous behavior.
We examine now the transitions in frequency that occur for an
initially asynchronous state. By adjusting the bifurcation parameter
to $\epsilon=0.5$ one insures the existence of an asymptotically stable
asynchronous state of the unperturbed system ($\lambda_1=0$)
~(\ref{eq.4})-(\ref{eq.6}). The frequency of the inertial oscillator
is chosen to be the same as defined above. We present results of
numerically calculated frequency ratios in Fig.~(\ref{fig:fig6}b).
Two curves are presented: the frequency ratio
$\Omega_e\omega_{tot}^{-1}$ between external driving frequency and
the frequency of the total extension $x_{tot}=\sum_{i=1}^N x_i$ versus $\Omega_e$
(squares) and the frequency ratio $\Omega_e\omega_{m}^{-1}$ versus $\Omega_e$ (dotted curve),
where $\omega_m$ is the spatially averaged frequency. One can observe that transitions from one collective
state to the other are controlled by changing the frequency
$\Omega_e$. In particular, there exists a $1:1$ plateau that
indicates the region of quasiperiodic asynchronous solutions.
 There are several transient regimes calculated for
different values of the external driving $\Omega_e$: asynchronous states for $\Omega_e=[0.2;1.2]$, 
the region of coexistence of synchronous and
\begin{figure}
\includegraphics[width=8.cm]{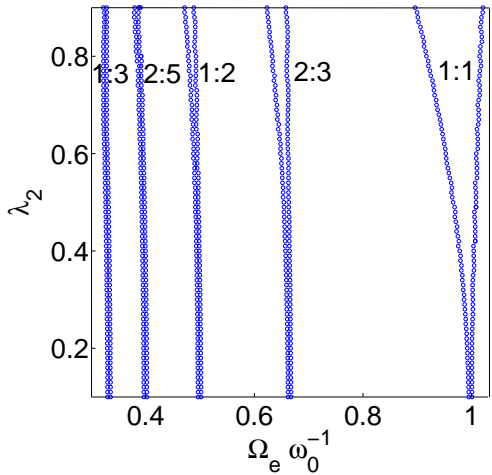}
\begin{center}
\caption{\label{fig:fig8} Arnol'd tongues: stability diagram for the
parameter space defined by amplitude $\lambda_2$ of external driving
and frequency ratio $\Omega_e\omega_0^{-1}$. The boundaries separating
stable limit cycles from unstable and quasiperiodic solutions are
shown. The frequency ratios of entrained limit cycle solutions
inside each region are indicated. Parameters
are:~$\epsilon=-0.2,\nu=0.2,\xi=0.3125,B=1,\omega=1,M=1,k=0.4 $.
Frequency of an unperturbed limit cycle solution $\omega_0=0.6337$.}
\end{center}
\end{figure}
asynchronous states for $\Omega_e=[1.2;1.35]$ at the border with the $2:1$
resonance. A large set of synchronized solutions emerges for the $2:1$
ratio when all the oscillatory elements are phase locked to the
$2:1$ frequency ratio regime. Note, that although for the initially
antisynchronous state the one--to--one stable entrainment regime is
not present (see Fig.~(\ref{fig:fig6}a)) this regime exist for the initially asynchronous state. For a 
detailed view on this regime we refer to Fig.~(\ref{fig:fig7}) which
shows a magnification of the $1:1$ plateau. Numerically calculated
nonlinear amplitude response (see Fig.~(\ref{fig:fig7}a)) for a
total chain extension $x_{tot}$ and the $1:1$ plateau in
Fig.~(\ref{fig:fig7}b) are magnified for ranges of frequencies
$\Omega_e$ that lie close to the $1:1$ resonance.  In
Fig.~(\ref{fig:fig7}c) the evolution of oscillator phases
$\phi_i=arg(z_i)$ is plotted for $N=50$. An example of quasiperiodic
solution for $1:1$ resonance is presented in
Fig.~(\ref{fig:fig7}d). Here, the oscillation of tension force $F$
is displayed versus the total chain extension $x_{tot}$.
\begin{figure}
\begin{center}
\includegraphics[width=8.cm]{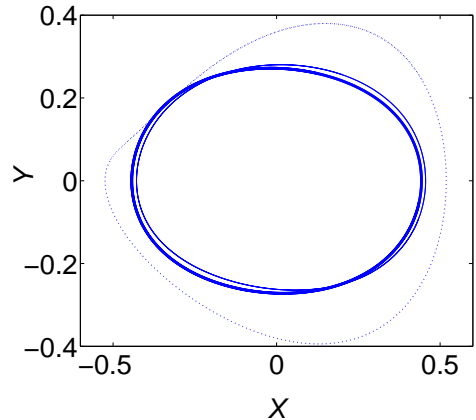}
\caption{\label{fig:fig9} Limit cycle solutions for the system~(\ref{eq.10})-(\ref{eq.11}). Thick line shows
the original limit cycle for the autonomous system~(\ref{eq.10})-(\ref{eq.11}) ($\lambda_2=0$). Two limit cycle
solutions (thin and dotted curves) are displayed for the system~(\ref{eq.10})-(\ref{eq.11})
with $\lambda_2=\omega$ and external frequency $\Omega_e=0.6337$
(dotted line), and with $\lambda_2=0.1 \omega$ (thin line).
Parameters are the same as in Fig.~(\ref{fig:fig8}).}
\end{center}
\end{figure}
Our results suggests, that the final collective behavior depends on
 the choice of initial collective state of the system. Indeed, as we have seen in Fig.~(\ref{fig:fig6}) the same
 stimulation frequency results in two distinct final states.
One interesting detail, that the stability diagrams indicate the largest set of the $2:1$
entrained solutions. The emergence of these solutions is due to the presence of
the multiplicative periodic modulation that produces the $2:1$ harmonic terms in the system
equations.
\section{System with parametric modulation of frequency of limit
cycle oscillators}\label{sec.4}
In this section we discuss the synchronization dynamics for the
system of self--oscillators with a parametric modulation of the frequency of self--oscillatory element. 
We modify a set of equations
~(\ref{eq.4})--(\ref{eq.6}) by setting $\lambda_1=0$ and introducing a periodic term
$g(\lambda_2,t)=\lambda_2 \cos \Omega_e t$ with the frequency $\Omega_e$ and the amplitude $\lambda_2$. 
The resulting equation for the $i$th oscillator can be written as follows:
\begin{equation}
\dot{z}_i=(\mathrm{i} \varrho(\lambda_2,t)+\epsilon)
z_i-B|z_i|^2z_i+F\xi^{-1},\label{eq.10}
\end{equation}
Where $\varrho(\lambda_2,t)=\omega+g(\lambda_2,t)$ is the time--dependent
frequency modulation term. Consequently, the
equation for the force yields:
\begin{eqnarray}
\dot{F}=-\frac{\xi F}{N M}+\frac{\xi}{N}\sum_{i=1}^N\mathrm{Re}\bigg\{\frac{d \left( B\vert z_i \vert^2
z_i+\mathrm{i} \varrho(\lambda_2,t) z_i\right)}{dt}\nonumber\\
-(\epsilon+\nu)
\dot{z}_i-\Omega_0^2 z_i\bigg\},\label{eq.11}
\end{eqnarray}
\subsection{Stability diagram for the synchronized state}
\label{sec.4.1}
 Analogously to the previous section where we first discussed the entrainment behavior
for the low--dimensional system, here we study first the case
of $N=1$ oscillator. By choosing $\epsilon<0$ and the linear oscillator
frequency $\Omega_0$ below the critical value $\Omega_c$ we assert
that the stable globally synchronized state exist for the autonomous system
(\ref{eq.10})-(\ref{eq.11})~($\lambda_2=0$). We intergate the set of equations~(\ref{eq.10})-(\ref{eq.11}) 
with initial conditions picked randomly near to the trivial solution. We carry out numerical stability
for a non--zero parametric modulation $g(\lambda_2,t)$.
In order to reach the limit of a strong excitation we define the
amplitude of the modulating term to be comparable with the amplitude
of a self--oscillating element frequency $\lambda_2/\omega\leq 1$. Figure~(\ref{fig:fig8}) shows
Arnol'd tongues calculated numerically. Inside the stability regions the frequency ratios 
are identified for each resonance. Examples of
the limit cycle solutions for the unperturbed system
($\lambda_2=0$) and for the perturbed system with $\lambda_2
\omega^{-1}=0.1$ and $\lambda_2\omega^{-1}=1$ are shown in
Fig.~(\ref{fig:fig9}).
\subsection{Amplitude--frequency response}
\label{sec.4.2}
\begin{figure}
 \begin{center}
 \includegraphics[width=9.cm]{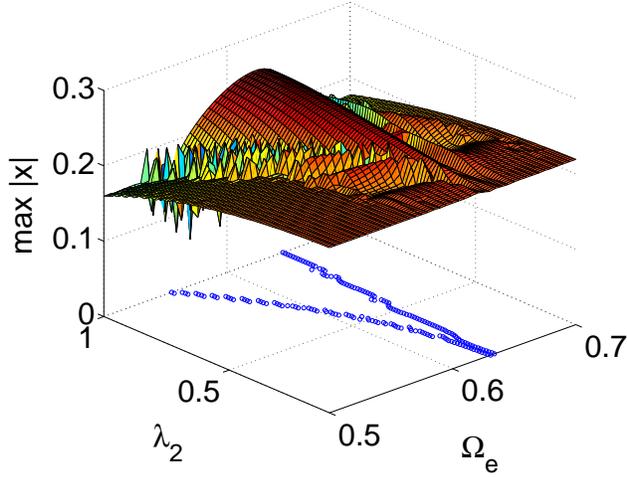}
 \caption{\label{fig:fig10} Amplitude--frequency response surface. Numerically calculated maximum of
 solution $x(t)$:~$\max|x|$ for the system (\ref{eq.10})-(\ref{eq.11}).
 The stability boundary of $1:1$ entrainment region is shown on the $(\Omega_e, \lambda_2)$ plane.
 Parameters: $\epsilon=-0.2,\nu=0.2,k=0.4, M=1,\xi=0.3125$. The frequency of an unperturbed limit cycle
is $\omega_0=0.6337$.}
 \end{center}
 \end{figure}
In Fig.~(\ref{fig:fig10}) the amplitude--response surface is
presented from numerical calculations of the maximum of solution $x(t)$ for the system~(\ref{eq.10})-(\ref{eq.11}). The maximum is estimated over the entire interval of
integration and obtained from the zero--averaged solution
$\max|x|=\max_{200 \pi\le t\le 300 \pi}|x(t)-\bar{x}|$. The borders of
stability region are displayed on the $(\Omega_e,\lambda_2)$ plane.
The response is smooth inside the one--to--one entrainment stability zone. It strongly shows nonlinear features for large
amplitude of the periodic driving: $\lambda_2\leq \omega$.
\subsection{Entrainment for inhomogeneous states} \label{sec.4.3}
\begin{figure}
\begin{center}
\includegraphics[width=8.cm]{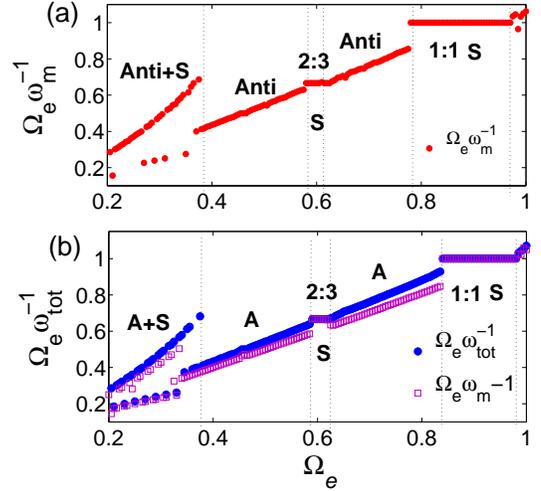}
\caption{\label{fig:fig11} Frequency ratio curves versus external
frequency $\Omega_e$ calculated from the Eqs.~(\ref{eq.10}) and
~(\ref{eq.11}). Parameters: $(\epsilon=0.1)$ for antisynchronous
state in (a), $(\epsilon=0.5)$ for asynchronous state in (b),
$\Omega_0=0.85$ and $\lambda_2=0.5 \omega$.}
\end{center}
\end{figure}
\begin{figure}
\begin{center}
\includegraphics[width=8.cm]{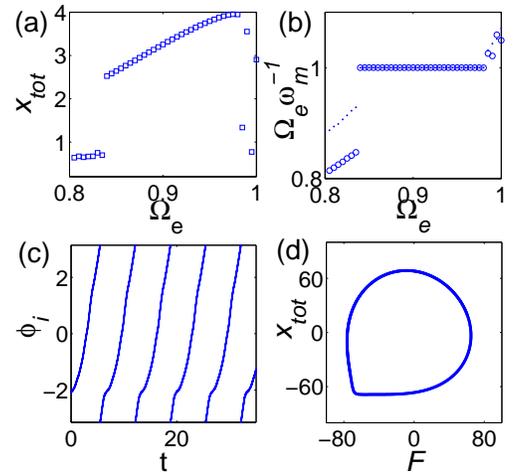}
\caption{\label{fig:fig12} Synchronized state of oscillators inside the
$1:1$ resonant region in Fig.~(\ref{fig:fig11}b). External driving
frequency $\Omega_e=0.85$ for~(c) and~(d).}
\end{center}
\end{figure}
Now we turn to the case of entrainment behavior for inhomogeneous
states. We compute frequency ratio curves for initially
inhomogeneous states of a chain of $N=100$. The stimulation frequency is chosen within the interval $[0.2;1]$
in order to include the $1:1$ resonance and the lower ratio resonances.
The results are reproduced in Fig.~(\ref{fig:fig11}) for an initially
antisynchronous state (a) and for an initially asynchronous state (b).
In both cases, one can distinguish two well--pronounced extended
plateaus at the $2:3$ and the $1:1$ resonances. The plateaus indicate the perfectly entrained solutions
with all the oscillatory units moving in phase and with the
frequency of $\Omega_e$. The remaining solutions that does not belong to the entrainment
regions are inhomogeneous states. They display
qualitatively different spacial dynamics. Each state is characterized by the individual oscillators
forming macroscopic clusters. The motion of the
individual units in a cluster is quasiperiodic. One can observe a region of
multistability for low values of $\Omega_e$. Stable coexisting
homogeneous and inhomogeneous states can be found inside this
region.
 In the following figure~(\ref{fig:fig12}) the magnification on the one--to--one synchronized
state is displayed. This state is obtain from an initially randomly distributed phases of individual elements in the
chain. Th amplitude of modulation is $\lambda_2=0.5 \omega$ and the external frequency $\Omega_e=0.85$.  The space--time plot 
of the synchronized state ca be also viewed in
Fig.~(\ref{fig:fig5}c). After a 
short transient all the oscillators begin to move in phase and with frequency equal to
$\Omega_e$. For the synchronized
state shown in Fig.~(\ref{fig:fig12}) the frequency of the total
extension $x_{tot}$ and the frequencies of self--oscillatory
elements math the stimulation frequency $\Omega_e$. In
Fig.~(\ref{fig:fig12}a) and (b) the amplitude response and the
frequency ratio are plotted versus $\Omega_e$. The periodic orbit in
the space of variables $F$ and $x_{tot}$ is also shown. 
\section{Discussion}\label{sec.5}
In this work we study entrainment behavior in an ensemble of
identical limit cycle oscillators coupled via geometrical constraint
and via a common force to an inertial linear oscillator.  The
external force due to a mass--spring load provides a global
coupling: it acts on each element of the ensemble. We consider an
external parametric modulation as a time--dependent force acting in
addition to the applied force due to a mass load  The
collective dynamics of oscillatory elements undergoes several
transitions between different inhomogeneous states. Such transitions
are ruled by the frequency and the type of parametric modulations
introduced in the model. Based on our description we find families
of stable oscillatory synchronized solutions corresponding to the
resonant external stimulation of the system. We discuss an influence
of two types of parametric excitations: periodic modulation of the
stiffness of the inertial load and modulation of the frequency of
self--oscillatory element. In particular, we demonstrate that for
the spatially synchronized state the second type of excitation leads
to larger stability domains than the first type of excitation.
Furthermore, we show that with the appropriate parametric modulation
introduced an initially inhomogeneous state of the system is driven
into the $1:1$ synchronized state with the frequency of collective
oscillations equal to the external frequency. Our conclusion is that
such globally synchronized behavior can not be achieved from an
initially inhomogeneous state by the periodic modulation of the
stiffness coefficient. Instead a new type of dynamical state is observed. The emergent spatially inhomogeneous
behavior can be described as a partially synchronized state: while
every individual element moves with a frequency higher than the
frequency of the total chain extension, the sum of displacements of all the elements
oscillate at a frequency equal to the external stimulation
frequency.

The amplitude--frequency characteristic shows a typical nonlinear
response behavior with a fold in the parameter space defined by
the driving frequency and the amplitude of the response. Our
numerical results show a good agreement with the quasiperiodic
approximation inside the one--to--one stable entrainment zone.

 The phenomena
discussed here have relevance and applications in nature and
laboratories. In particular, the autonomous model and its
time--dependent analogue provide a phenomenological approach to
model oscillatory dynamics of an active muscle fiber. Experimental
\textit{in vitro} observations on skinned muscle fiber show
evidence of different oscillatory regimes generated by spontaneous
contractions of muscle elementary units,
sarcomeres~\cite{yasuda1996}. Depending on the conditions of the
experiment, oscillatory contractions can undergo a synchronized
activity~\cite{yasuda1996} or a spatially inhomogeneous
(asynchronous or antisynchronous)
activity~\cite{okamura1988,fukuda1996}. In experiments the
signalling between adjacent sarcomeres is chemically regulated. The
state of activation of one sarcomere is affected by the state of the
adjacent sarcomere~\cite{yasuda1996}.
Our current model does not consider the effect
of coupling between adjacent sites in the chain of elements.
In the future we plan to include local coupling between
elements and to consider its effect on spatiotemporal dynamics. Detailed
study of the corresponding regimes of spatiotemporal behavior is left to future work.

Experimental work and results from theoretical analysis of our
phenomenological model promise new interesting insights into
collective behavior of oscillations in isolated fibrillar muscle.
Periodic modulation of the model parameters considered here might
serve as an appropriate setting for modeling active muscle
contractions under a specially designed external mechanical and
chemical stimulation.
\section{Appendix}\label{sec.6}
In the appendix, we discuss a quasiperiodic
approximation~\cite{bogoliubov1961} used for derivation of the
nonlinear amplitude--frequency dependence. The derivation is carried
out for the first type of parametric modulation discussed in the
paper. It can be seen from Fig.~(\ref{fig:fig2}) that the original
unperturbed limit cycle is close to the ellipse centered at zero.
Solutions for a non--zero small perturbation $\lambda_1\ll k$ are
nearly close to the disturbed elliptical form with their centers
shifted away from zero. These observations lead us to seek the
solutions expansion in form of a finite series of harmonic terms,
where the largest contribution terms represent 
solution for an ellipse. The system ~(\ref{eq.4})-(\ref{eq.6}) can
be rewritten in a real form after introducing $x=\mathrm{Re}(z)$ and
$y=\mathrm{Im}(z)$ variables with $x$ a displacement of the
oscillator and $y$ a time--dependent variable nonlinearly coupled to
$x$:
\begin{subequations}
\label{eq.12}
\begin{eqnarray}
M/\xi\ddot{x}&=&-\left(1+M \nu/\xi\right)\dot{x}- K(\lambda_1,t)/\xi
x\nonumber\\&&-\omega y+\epsilon x
-B(x^2+y^2)x,\\
\dot{y}&=&\omega x+\epsilon y-B(x^2+y^2)y,\end{eqnarray}
\end{subequations}
 For the non--autonomous system (\ref{eq.12}) a limit cycle
solution can be written in the quasiperiodic form~\cite{tomita1977}:
\begin{subequations}
\label{eq.13}
\begin{eqnarray}
x&=&a_1 \cos~\Omega t+b_1 \sin~\Omega t+a_0+\ldots,\\
y&=&b_2 \sin~\Omega t+a_2 \cos~\Omega t+a_g+\ldots.
\end{eqnarray}
\end{subequations}
Where $\Omega$ is the frequency of a new limit cycle for the time--dependent
system~(\ref{eq.12}). The amplitude coefficients in the
expansion ~(\ref{eq.13}) depend on the size of the original unperturbed
limit cycle solution and are ordered according to the powers of
$\epsilon$: $a_1, b_2,a_g$ are of order $O(\epsilon^{1/2})$,~ $b_1,a_2$ are of order
$O(\epsilon)$ and $a_0=O(\epsilon^{3/2})$. If the
driving frequency $\Omega_e$ is near to the frequency $\omega_0$
of the limit cycle solution for the unperturbed system
$(\lambda_1=0)$ in~(\ref{eq.4})-(\ref{eq.6}) a reasonable approximation for the $1:1$
lock--in solution can be made by retaining only terms to the accuracy of $O(\epsilon^2)$
 in Eqs.~(\ref{eq.13}). We would not consider the
terms of order equal to $O(\epsilon^2)$ in the
approximation presented here. Coefficients of the
expansion~(\ref{eq.13}) can be expressed via coefficients
in~(\ref{eq.7}) as follows:
\begin{subequations}
\label{eq.14}
\begin{eqnarray}
a_1&=&\frac{c_1+c_{-1}}{2},~b_1=\frac{\mathrm{i} (c_{-1}-c_1)}{2},~a_0=\bar{x},\\
a_2&=&\frac{d_1+d_{-1}}{2},~b_2=\frac{\mathrm{i} (d_{-1}-d_1)}{2},~a_g=\bar{y}.
\end{eqnarray}
\end{subequations}
We introduce a small parameter $\Delta=\Omega-\Omega_e=
O(\epsilon^2)$ to be a frequency mismatch between the external
frequency and the resulting limit cycle frequency. Let us define a linear frequency:
\begin{equation}
\Omega_0=\omega/(1+M\nu/\xi)^{1/2}.\nonumber
\end{equation}
 Consequently, one finds that the frequency
ratio $\Omega_e\omega_0^{-1}$ can be expressed as follows:
\begin{equation}
\Omega_e\omega_0^{-1}\approx\Omega\Omega_0^{-1}(1-\Delta).\nonumber
\end{equation}
 In the next
step, we use the quasiperiodic approximation~(\ref{eq.13}) to derive
a final form of the amplitude--frequency response. We assume that the following expansion
holds true: $k=k_0+\epsilon k_1$, with
$k_0=M\Omega_0^2$. For the sake of simplicity, we introduce the rescaling of the parameters:
\begin{equation}
\tilde{m}\rightarrow M/\xi, \tilde{k}\rightarrow k/\xi,~\tilde{\lambda_1}\rightarrow\lambda_1/\xi.\nonumber
\end{equation}
Upon substituting the quasiperiodic form~(\ref{eq.13}) into Eq.~(\ref{eq.5}) and collecting various terms according
to the order of harmonics we obtain a set of nonlinear equations:
\begin{subequations}
\label{eq.15}
\begin{eqnarray}
\tilde{m} \ddot{a_0}=&&-(1+\tilde{m} \nu)\dot{ a_0}+(\epsilon-\tilde{k})a_0\nonumber\\&&
-\omega a_g-\tilde{\lambda_1} a_1/2-g_1 a_0\nonumber\\&&-a_1 g_2-b_1 g_3,\\
\dot{a_g}=&&\epsilon a_g+\omega a_0-a_g g_1-a_2 g_2\nonumber\\&&-b_2 g_3,\\
\ddot{a_1}=&&-(1+\tilde{m} \nu)\dot{a_1}+\Omega^2 \tilde{m} a_1\nonumber\\&&-(1+\tilde{m}\nu)\Omega b_1
+(\epsilon-\tilde{k}) a_1-\omega a_2\nonumber\\&&-\tilde{\lambda_1} a_0-a_1 g_1
-2 a_0 g_2\nonumber\\&&-a_1 g_4/4-b_1 g_5/2,\\
\dot{a_2}=&&-\Omega b_2+\epsilon a_2+\omega a_1-2 a_g  g_2\nonumber\\&&-a_2 g_1-a_2 g_4/4-b_2 g_5/2,\nonumber\\
\dot{b_2}=&&\Omega a_2+\epsilon b_2+\omega b_1-2 a_g g_3-b_2 g_1\nonumber\\&&-a_2 g_5/2+b_2 g_4/4,\\
 \tilde{m}\ddot{b_1}=&&-(1+\tilde{m}\nu) \dot{b_1}+\tilde{m}\Omega^2 b_1\nonumber\\&&+(1+\tilde{m}\nu)\Omega a_1
 +(\epsilon-\tilde{k})b_1-\omega b_2\nonumber\\&&-2 a_0 g_3-a_1 g_5/2-b_1 g_1\nonumber\\&&-b_1 g_4/4,
\end{eqnarray}
\end{subequations}
where $g_1,g_2,g_3$ are nonlinear functions of the coefficients
in~(\ref{eq.13}):
\begin{eqnarray}
g_1&=&B\bigg(a_0^2+a_g^2+\frac{a_1^2}{2}+\frac{a_2^2}{2}+\frac{b_2^2}{2}+\frac{b_1^2}{2}\bigg),\nonumber\\
g_2&=&B(a_0 a_1+a_g a_2),\nonumber\\
g_3&=&B(a_0 b_1+a_g b_2),\nonumber\\
g_4&=&B(a_1^2-b_1^2+a_2^2-b_2^2),\nonumber\\
g_5&=&B(a_1 b_1+a_2 b_2).\nonumber\\
\end{eqnarray}
We would not consider the
terms of order $O(\epsilon^{3/2})$ in the approximation presented here.
Let us make the assumption that for the entrained limit cycle solution~(\ref{eq.13})
the coefficients in the expansion~(\ref{eq.13}) are constant or vary slowly with time.
Then one can neglect derivatives from the equations for the coefficients~(\ref{eq.15}).
By neglecting  terms of order  $O(\epsilon^{3/2})$
one can express the coefficients $a_2,a_g,b_1$ and $b_2$:
\begin{subequations}
\label{eq.16}
\begin{eqnarray}
a_g=-\frac{\tilde{\lambda}_1}{2 \omega}a_1,\\
a_2=-\frac{\omega}{\Omega} b_1,~b_2=\frac{\omega}{\Omega}a_1,\\
b_1=\frac{(1+\tilde{m}\nu)\Omega-\omega^2/\Omega}{\tilde{m}(\Omega_0^2-\Omega^2)}a_1.
\end{eqnarray}
\end{subequations}
Let us use the substitution $\rho=a_1^2$. Our final goal is to obtain a family of 
amplitude--frequency curves:
\begin{equation}
 \Gamma_{\tilde{\lambda}_1}(\Omega_e):\Omega_e\omega_0^{-1}\rightarrow
\rho(\Omega \Omega_0^{-1}).\nonumber
\end{equation}
  After substituting the expressions~(\ref{eq.16}) for the coefficients into Eqs.~(\ref{eq.15}) and eliminating
terms of the order $O(\epsilon^{3/2})$ we write down the approximate amplitude--frequency relation: 
\begin{widetext}
\begin{eqnarray}
a_0\tilde{\lambda}_1^2=&&\frac{B^2}{16}\bigg[\tilde{\lambda}_1^2+\left(3+\frac{\omega^2}{\Omega^2}\right)\left(1+\frac{1}{\tilde{m}^2}
\frac{\omega^4}{\Omega_0^4\Omega^2}\right)\bigg]^2\rho^3-\frac{B}{2}\bigg[\left(\tilde{m}-\frac{\omega^4}{\tilde{m} \Omega_0^4 \Omega^2}\right)(\Omega^2-\Omega_0^2)+\epsilon\left(1-\tilde{k}_1\right)\bigg]\times\nonumber\\&&\bigg(3+\frac{\omega^2}{\Omega^2}\bigg)\Bigg(1+\frac{1}{\tilde{m}^2\Omega_0^4\Omega^2}\bigg)\rho^2
 +\bigg[\bigg(\tilde{m}^2+\bigg(\frac{\omega^4}{\tilde{m}
\Omega_0^6}\bigg)^2\bigg)\left(\Omega^2-\Omega_0^2\right)^2-2\bigg(\frac{\omega^2}{\Omega}
-(1+\tilde{m}\nu)\Omega\bigg)^2+2\epsilon(\tilde{k}_1-1)\times\cr
&&\bigg(\tilde{m}-\frac{\omega^4}{\tilde{m}\Omega_0^4\Omega^2}\bigg)(\Omega_0^2-\Omega^2)\bigg]\rho,\label{eq.17}
\end{eqnarray}
\end{widetext}
 Finally, what remains to be done is to determine a set of parameters
for calculating the relation~(\ref{eq.17}).
Let us define the parameters in order to compare with the numerical results:
$\epsilon=-0.2,\nu=0.2,M=1,\omega=1,\xi=0.3125,
\Omega_0=0.7861,\lambda_1=k_0, a_0=0.01 \lambda_1^{-2}$. We have used
$\Delta=0.07~\Omega_0\Omega^{-1},~k_1=0.4 \xi$ in order to obtain
a reasonable fitting to the numerical amplitude--frequency response
in Fig.~(\ref{fig:fig4}).
\section{Acknowlegements}\label{sec.7}
 The author thanks E. Nicola for many useful comments and
suggestions regarding implementation of numerical code. The author
is grateful to S. Ares for his comments on the manuscript. 
The author would like to acknowledge M. Zapotocky for helpful
discussions. This research was partially funded by the
VolkswagenStiftung Foundation and the Max-Planck-Gesellschaft.




\begin{thebibliography}{10}
\bibitem{glass1991}
L. Glass, A. Shrier, in L. Glass, P. Hunter, A. McCulloch,(Eds.),Theory of Heart,Spinger, New York,289 (1991)
\bibitem{graves1986}
 C. Graves, L. Glass, D. Laporta, R. Meloche, A. Grassino, Am. J. Physiol.(Regulat. Integrat. Comp. Physiol. 19) 250 R902 (1986)
\bibitem{dickinson2006}
M. H. Dickinson, Current Biology 16 R309 (2006)
\bibitem{nalbach1988}
 G. Nalbach, Proc. {G}{\"o}ttingen {N}eurobiol. {C}onf. 16 131 (1988)
\bibitem{lehmann1997}
 F. -O Lehmann, M. H. Dickinson, J. Exp. Biol. 200 1133 (1997)
\bibitem{pikovsky2001}
 A. Pikovsky, M. Rosenblum, J. Kurths, Synchronization: a Universal Concept in Nonlinear Science,Cambridge University Press, New York,(2001)
\bibitem{glass2001}
 L. Glass, Nature 401 277 (2001)
\bibitem{abarbanel2006}
H. D. Abarbanel, M. I. Rabinovich, Current Opinion in Neurobiology 11 423 2006
\bibitem{chate1999}
 H. Chate, A. Pikovsky, O. Rudzick, Physica D  131 17 (1999)
\bibitem{osipov2007}
G. V. Osipov, J. Kurths, C. Zhou, Synchronization in Oscillatory Networks, Springer-Verlag, Berlin Heidelberg, (2007)
\bibitem{afraimovich1994}
 V. S. Afraimovich, V. I. Nekorkin, G. V. Osipov, V. D. Shalfeev, Stability, Structures and Chaos in Nonlinear Synchronization Networks,
 World Scientific, Singapure, (1994)
\bibitem{rudiger2007}
 S. R{\"u}diger, E. M. Nicola, J. Casademunt, L. Kramer, Phys. Rep. 447 73 (2007)
\bibitem{battogtokh1996}
 D. Battogtokh, A. Mikhailov, Physica D 90 84 (1996)
\bibitem{kuramoto1984}
 Y. Kuramoto, Chemical Oscillations, Waves and Turbulence, Springer-Verlag, Berlin,(1984)
\bibitem{coullet1992}
 P. Coullet, K. Emilsson, Physica D 61 119 (1992)
\bibitem{drendel1984}
 S. D. Drendel, N. P. Hors, V. A. Vasiliev, Dynamics of Cell Populations, Nizhny Novgorod University Press, Nizhny Novgorod, (1984)(in Russian)
\bibitem{vadivasova2001}
T. E. Vadivasova, G. E. Strelkova, V. S. Anishchenko, Phys. Rev. E  63 036225 (2001)
\bibitem{hakim1992}
 H. Hakim, W.-J. Rappel, Phys. Rev. A 46 7347 (1992)
\bibitem{topaj2002}
D. Topaj, A. Pikovsky, Physica D 170 118 (2002)
\bibitem{vilfan2003}
 A. Vilfan, T. Duke, Phys. Rev. Lett. 91 114101 (2003)
\bibitem{yasuda1996}
 K. Yasuda, Y. Shindo, S. Ishiwata, Biophys. J. 70 1823 (1996)
\bibitem{laskar1992},
 J. Laskar, C. Froeschl{\'e}, A. Celletti, Physica D 56 253 (1992)
\bibitem{daido1996}
 H. Daido H, Physica D 91 24 (1996)
\bibitem{ermentrout1991}
G. B. Ermentrout, N. Kopell, J. Math. Biol. 29 195 (1991)
\bibitem{fukuda1996}
 N. Fukuda, H. Fujita, T. Fujita and S. Ishiwata, Eur. J. Physiol. 433 1 (1996)
\bibitem{okamura1988},
 N. Okamura, S. Ishiwata, J. Muscle Res.Cell Motil. 9 111  (1988)
\bibitem{bogoliubov1961},
 N. N. Bogoliubov, Y. A. Mitropolsky, Asymptotic Methods in the Theory of Nonlinear Oscillators, Gordon and Breach,
 New York,(1961)
\bibitem{tomita1977}
  K. Tomita, T. Kai, F. Hikami, Prog. Theor. Phys.  57 1159 (1997)
\end{thebibliography}
\end{document}